\documentclass{article}

\usepackage{caption}
\usepackage{subcaption}
\usepackage{amssymb}
\usepackage{graphicx}
\usepackage{dcolumn}
\usepackage{bm}
\usepackage{amsmath}

\begin{document}

\title{\bf Gravity's rainbow: a bridge between LQC and DSR}
\author{M. A. Gorji$^{1}$\thanks{email: m.gorji@stu.umz.ac.ir},
\hspace{.2 cm} K. Nozari$^1$\thanks{{email: knozari@umz.ac.ir}}
\hspace{.2cm} and \hspace{.2cm}B. Vakili$^2$\thanks{email:
b.vakili@iauctb.ac.ir}\vspace{.2cm}\\ $^1${\small {\it Department of
Physics, Faculty of Basic Sciences, University of Mazandaran},}
\\{\small {\it P.O. Box 47416-95447, Babolsar, Iran}}\\$^2${\small
{\it Department of Physics, Central Tehran Branch, Islamic Azad
University, Tehran, Iran}}}\maketitle

\begin{abstract}
The doubly special relativity (DSR) theories are constructed in order
to take into account an observer-independent length scale in special
relativity framework. It is widely believed that any quantum theory
of gravity would reduce to a DSR model at the flat limit when purely
gravitational and quantum mechanical effects are negligible. Gravity's
rainbow is a simple generalization of DSR theories to incorporate
gravity. In this paper, we show that the effective Friedmann equations
that are suggested by loop quantum cosmology (LQC) can be exactly
reobtained in rainbow cosmology setup. The deformed geometry of LQC
then fixes the modified dispersion relation and results in a
unique DSR model. In comparison with standard LQC scenario where only
the geometry is modified, both geometry and matter parts get
modified in our setup. In this respect, we show that the total
number of microstates for the universe is finite which suggests the
statistical origin of the energy and entropy density bounds. These
results explicitly show that the DSR theories are appropriate
candidates for the flat limit of loop quantum gravity.
\vspace{5mm}\\
PACS numbers: 04.60.Bc\\
Keywords: Phenomenology of Quantum Gravity, Doubly Special Relativity
\end{abstract}

\section{Introduction}
In the standard model of cosmology, the universe starts from the big
bang singularity where the classical Einstein's equations are no longer
applicable. The Hubble parameter diverges when the size of the universe
approaches zero at the singularity. Beside, the singularity is also
consistently understandable from the matter part where the standard
statistical mechanics formalism applies to the early radiation
dominated universe: The energy density of the universe,
determined by the standard Stefan-Boltzmann law, and the associated
entropy density diverge at big bang singularity. The energy density is
directly related to the Hubble parameter in standard Friedmann
cosmology while the entropy density is proportional to the inverse of
the size of the universe through the adiabatic evolution of the
universe. Therefore, the divergences of the geometrical and
thermodynamical quantities are consistent in standard early universe
cosmology. But the big bang singularity remains as an unsolved
problem: Both of the classical Einstein's equations and standard
thermostatistics cannot explain the state of the universe at such a high
energy regime. Quantum gravity candidates such as loop quantum gravity
and string theory suggest the existence of a minimum length scale (of
the order of the Planck length) below of which no other length can be
probed \cite{Loop,String}. The existence of this minimum length
scale evidently prevents the big bang singularity in loop quantum
cosmology (LQC) \cite{LQC,LQC-BB} and string cosmology
\cite{String-Cos}. In LQC, the big bang singularity
problem can be resolved in such a way that the singularity replaces with a
quantum bounce \cite{LQC-BB}. The Hubble parameter gets a maximum and
there is a nonzero minimum size for the universe at the
bounce. Existence of this geometrical maximum bound on the Hubble
parameter immediately implies a maximum bound for the energy density
through the Friedmann equation. Also, a maximum entropy density bound
arises through the adiabatic evolution of the universe since the
universe has a nonzero minimum size in LQC. These bounds on
the energy and entropy densities, however, are not naturally supported
by the standard statistical mechanics formalism. This inconsistency is
understandable: The statistical mechanics formalism should be of
course modified in order to include quantum gravity (minimal length)
effects when one considers the modified geometry of LQC for the
geometric part of the Einstein's equations. The question then arises:
How the statistical mechanics would be modified in the context of LQC
while it only deals with the quantization of the geometry and not the
matter? To answer this question, we note that at the flat limit, with
which we are interested in the context of statistical mechanics, the
fundamental quantum theory of gravity such as loop quantum gravity
would reduce to a deformed (doubly) special relativity (DSR)
which supports the existence of minimum observer-independent length
scale \cite{DSR-FL}. In the absence of the final and well-established
quantum theory of gravity, one can do in reverse: starting from the
standard special relativity and deform it in such a way that it includes
an invariant length scale. This is the main idea of the DSR theories
\cite{DSR}. The statistical mechanics in DSR framework is widely studied
and, interestingly, it is shown that the entropy and energy densities
get maximum bounds in this setup \cite{DSR-THR}. In this respect, in
Ref. \cite{DSR-DoS2}, it is claimed that the modified geometry of LQC is
consistent with the statistical mechanics based on DSR and not the
standard special relativity (see also Ref. \cite{Ronco}). But this 
consistency has not been explicitly
confirmed. In this paper, we reobtain the effective Friedmann equation
of LQC in gravity's rainbow (doubly general relativity) formalism which
is a simple generalization of DSR theories to include gravity
\cite{rainbow}. This result explicitly shows the relevance of the DSR
theories for the flat limit of loop quantum gravity. Moreover, the
quantum gravity modifications to the statistical mechanics formalism
naturally arise in our setup. The maximum bounds on Hubble parameter,
energy density, and entropy density are then completely understandable
from both geometrical and thermodynamical point of views in our
setup.

\section{Gravity's Rainbow Cosmology}
In recent years, the idea of gravity's rainbow \cite{rainbow} attracted
some attentions specially in black hole physics \cite{rainbow-BH} and
cosmology \cite{rainbow-Cos}. In this section we briefly review the
early radiation dominated universe cosmology in this framework.

The DSR theories are generally defined by the deformed dispersion
relation\footnote{We work in units $\hbar=c_0=k_{_B}=1$, where
$\hbar$, $c_0$, and $k_{_B}$ are the Planck constant, {\it standard}
speed of light in vacuum, and Boltzmann constant respectively.}
\begin{equation}\label{MSR}
f^2(\lambda{E})\,E^2-g^2(\lambda{E})\,p^2=m^2\,,
\end{equation}
where $\lambda$ is the invariant UV scale which signals the existence
of a minimal observer-independent length scale in this setup. The
different DSR models are determined by the different functional forms
for $f(\lambda{E})$ and $g(\lambda{E})$. All of these
models respect the correspondence principle as $\lim_{\lambda{E}
\rightarrow0}f,g=1$ while they have different behaviors at UV regime
\cite{AdS}. The Lorentz symmetry breaks at UV regime for some models
\cite{DSR} or is preserved by nonlinear action of the Lorentz group on
the associated curved momentum space \cite{Glikman,DSR2,DSR-RL}. In
contrast to the standard special relativity, defining the position
space to be dual to the curved momentum spaces in DSR setup is highly
nontrivial \cite{DSR-PS0,DSR-PS}. In Ref. \cite{DSR-PS}, the authors
suggest a novel construction of the dual position space. It is claimed
that the metric on the dual position space will be $ds^2=-f^{-2}dt^2+
g^{-2}\delta_{ij} dx^idx^j$ in order to have plane wave solution for
the corresponding free field theory. The setup is then generalized to
the curved spacetime to incorporate gravity which resulted in
gravity's rainbow or doubly general relativity \cite{rainbow}. The
modified equivalence principle is then investigated that leads to the
one parameter family of connections and curvature tensors and
therefore one parameter family of Einstein's equations (see Ref.
\cite{rainbow} for more details)
\begin{equation}\label{E-E}
G_{\mu\nu}(\lambda{E})=8\pi{G(\lambda{E})}\,
T_{\mu\nu}(\lambda{E})\,,
\end{equation}
where $G(\lambda{E})$ is the effective gravitational coupling constant
at UV regime which should lead to the standard Newton's constant $G_0$
at low energy regime as $\lim_{\lambda{E}\rightarrow0}G=G_0$. For the
case of the flat FLRW spacetime, the corresponding rainbow metric is
given by \cite{rainbow}
\begin{equation}\label{Metric-Rainbow}
ds^2=f^{-2}\Big(-dt^2+a^2(t)\,\delta_{ij}dx^idx^j\Big)\,,
\end{equation}
where we have assumed $f=g$ in the relation (\ref{MSR}). This choice
leads to the constant speed of light through the modified dispersion
relation (\ref{MSR}) as $c(\lambda{E})=g/f=c_0=1$ (in our units). Note
that the DSR theories generally lead to the varying speed of light
\cite{DSR-Inflation,AdS}. Also the definition of speed of light in
DSR theories is not unique (see Refs. \cite{DSR-PS,DSR-Vel} for more
details). In gravity's rainbow formalism, the definition $c=g/f$
is more acceptable since it coincides with the speed of null signals
defined by the associated rainbow metric as $ds^2=0$. Thus, our choice
of $f=g$ has a physical consequence that the speed of light is
constant even in the UV regime similar to the well-known original
Magueijo-Smolin model \cite{DSR2}. We also assume that the
gravitational coupling constant being the standard Newton's constant
such that $G=G_0$ as one usually assumes in the context of gravity's
rainbow. These choices for the speed of light and gravitational
constant are reasonable since we would like to explore the relation
between the rainbow cosmology and LQC and both the speed
of light and gravitational coupling constant have their standard
constant definitions in LQC. The matter part is determined by
the perfect fluid energy-momentum tensor
\begin{equation}\label{EMT}
T_{\mu\nu}=\rho\,{u_{\mu}
u_{\nu}}+P\,(g_{\mu\nu}+u_{\mu}u_{\nu})\,,
\end{equation}
where $\rho$ and $P$ are the energy density and pressure. The four
velocity $u_\mu$ depends on $\lambda{E}$ as $u_\mu=(f^{-1},0,0,0)$
to respect the normalization relation $g^{\mu\nu}u_\mu{u_\nu}=-1$.

Substituting the metric (\ref{Metric-Rainbow}) together with the
energy-momentum tensor (\ref{EMT}) into the Einstein's equations
(\ref{E-E}), one finds the following deformed Friedmann and Raychaudhuri
equations (see Ref. \cite{DSR-Loop} for details)
\begin{equation}\label{Friedmann}
H^2-2H{\dot f}f^{-1}+{\dot f}^2f^{-2}=
\frac{\kappa}{3}\,f^{-2}\rho\,,
\end{equation}
\begin{equation}\label{Raychaudhuri}
{\dot H}+H{\dot f}f^{-1}+{\ddot f}f^{-1}=-
\frac{\kappa}{6}\,f^{-2}(\rho+P)\,,
\end{equation}
where $\kappa=8\pi{G_0}$. The deformed Bianchi identities, then give
the following modified energy conservation relation
\begin{equation}\label{Bianchi}
{\dot \rho}+3(H-{\dot f}f^{-1})(\rho+P)=0\,.
\end{equation}
Two of the three equations (\ref{Friedmann}), (\ref{Raychaudhuri}),
and (\ref{Bianchi}) are independent which completely determine the
dynamics of the universe in our setup.

\section{LQC versus DSR}
In relations (\ref{Friedmann}), (\ref{Raychaudhuri}) and
(\ref{Bianchi}), the evolution is considered with respect to the
cosmic time $t$, while $f$ depends on kinematical energy $E$ and is not
an explicit function of $t$. Also, the energy density $\rho$ and
pressure $P$ do not explicitly depend on $E$. More precisely, $E$ is
the total kinematical energy of the particles at which the spacetime
geometry is probed \cite{rainbow}. Thus, it depends on $t$ and
therefore $f$ depends on $t$ implicitly. On the other hand, for the case of early
radiation dominated universe, the energy density and pressure in
(\ref{EMT}) depend on temperature and therefore on $t$ through the
adiabatic condition. But, referring to Eq. (\ref{E-E}),
how these thermodynamical quantities may depend on the energy scale
$E$ at which the spacetime geometry is probed? The key is indeed the
modified dispersion relation (\ref{MSR}) which immediately leads to
the modification of the density of states
\cite{DSR-DoS2,DSR-Inflation,DSR-DoS}. For the case of massless
particles in our setup with $f=g$ and therefore $E=p$, the form of the
associated density of states remains unchanged \cite{DSR-DoS}. But it
is important to note that depending on the functional form of $f$, a
UV cutoff $E<\lambda^{-1}$ can arise which modifies the thermodynamical
quantities through the ranges of the integrals over the momenta (see
for instance Ref. \cite{DSR-DoS2}). Therefore, the energy-momentum
tensor depends on $\lambda{E}$ as it is shown in (\ref{E-E}). In
summary, at least for the radiation dominated universe, the bridge
between the kinematical energy of particles $E$ and the energy density
$\rho$ is given by the temperature when one reasonably identifies the
kinematical energy $E$ with the thermodynamical internal energy $U$
that is the statistical average of $E$ \cite{DSR-Loop}. More precisely,
this relation is indeed between the dimensionless quantities $\lambda{
E}$ and $\frac{\rho}{\rho_{\max}}$. We therefore assume
\begin{equation}\label{equivalence}
\lambda{E}=\frac{\rho}{\rho_{\max}}\,,
\end{equation}
where $\rho_{\max}$ is the maximum value of the energy density
corresponding to the maximal kinematical energy $E=\lambda^{-1}$. This
relation will be fixed explicitly in section 5 through
statistical considerations.

Let us rewrite the Friedmann equation (\ref{Friedmann}) and the
energy conservation relation (\ref{Bianchi}) in more appropriate forms
\begin{equation}\label{Friedmann2}
H^2=\frac{\kappa}{3}\,f^{-2}\rho+2H{\dot f}
f^{-1}-{\dot f}^2f^{-2}\,,
\end{equation}
\begin{equation}\label{Bianchi2}
{\dot \rho}=-3\big(H-{\dot f}f^{-1}\big)\big(1+w(\rho)\big)\rho\,,
\end{equation}
where we have substituted $w(\rho)=P(\rho)/\rho$. Note that the
equation of state parameter $w$ is modified in rainbow cosmology since
the density of states gets modified through the cutoff $E<
\lambda^{-1}$ in this setup \cite{DSR-DoS2,DSR-Inflation} (see also
section 5 of the present paper). At the low energy limit ($\lambda{E}\rightarrow0$), $f
\rightarrow1$ and equations (\ref{Friedmann2}) and (\ref{Bianchi2})
reduce to their standard counterparts. The functional form of $f$ in
(\ref{Friedmann2}) and (\ref{Bianchi2}) is however not explicitly
fixed and, therefore, there are many Friedmann equations for the
different choices of $f$ (see for instance Refs.
\cite{DSR-DoS2,rainbow,DSR-Loop}). There is no clear reason to
prefer one set of equations from the others and it is natural to
expect that the ultimate quantum theory of gravity finally fixes the
functional form of $f$. In the absence of such a theory, we can refer to
quantum gravity candidates such as loop quantum gravity and
string theory. Indeed, the (fixed) deformed Friedmann equations are
investigated in the frameworks of LQC \cite{LQC-BB} and string
cosmology \cite{String-Cos}. Taking the equivalence relation
(\ref{equivalence}) into account, we can fix the functional form of
$f$ by referring to these candidates. This is what we can do at least in the absence
of full quantum gravity theory. Here we would like to fix $f$ by
invoking the effective Friedmann equations that are suggested by LQC.
Following the method we introduce in this paper, one can also
fix the functional form of $f$ through the string cosmology setup
\cite{String-Cos}.

For the case of flat FLRW universe, the modified Friedmann equation
in LQC framework is given by \cite{Ashtekar-EDB}
\begin{equation}\label{LQC-Friedmann}
H^2=\frac{\kappa}{3}\,\rho\Big(1-\frac{\rho}{\rho_c}
\Big)\,,
\end{equation}
where
\begin{equation}\label{rho-c}
\rho_c=\rho_{\max}=\frac{3}{8{\pi}G_0\alpha_0\gamma^2
l_{_{\rm Pl}}^2}\,,
\end{equation}
is the critical energy density which is also the maximum accessible
energy density for the universe. This quantity coincides with the maximum
energy density $\rho_{\max}$ that perviously has been defined in
(\ref{equivalence}). The two dimensionless parameters $\gamma$ and
$\alpha_0$, are the Barbero-Immirzi parameter and the minimum
eigenvalue of the area operator \cite{Ashtekar-EDB} respectively. The
Barbero-Immirzi parameter is fixed as $\gamma=0.2375$ through the
black hole entropy calculation \cite{BI} and the minimum area
eigenvalue is $\alpha_0=4\sqrt{3}\pi\gamma=5.166$ for the homogenous
and isotropic models. Both of these parameters disappear at the
classical limit $\rho\ll\rho_{\max}=0.41\rho_{_{\rm Pl}}$ where
(\ref{LQC-Friedmann}) coincides with the standard Friedmann equation.
The energy conservation relation remains unchanged in LQC setup and
for the case of radiation dominated universe is given by
\begin{equation}\label{LQC-ECR}
{\dot \rho}+4H\rho=0\,.
\end{equation}
The modified Friedmann equation (\ref{LQC-Friedmann}) is obtained from
the unique representation of holonomy-flux algebra in LQC.
We use this unique equation to fix the from of $f$ in the
corresponding Friedmann equation (\ref{Friedmann2})
suggested by gravity's rainbow formalism. To do this end, we use the chain
rule ${\dot f}={\dot\rho}\frac{df}{d\rho}=-4H\rho\frac{df}{d\rho}$
where we have used (\ref{LQC-ECR}). Substituting this into the right
hand side of (\ref{Friedmann2}) and then substituting $H$ from
(\ref{LQC-Friedmann}) into the left hand side, after some
manipulations, we obtain the following first order differential
equation
\begin{equation}\label{DE}
4\rho\frac{df}{d\rho}+f=\Big(1-\frac{\rho
}{\rho_{\max}}\Big)^{-\frac{1}{2}}\,,
\end{equation}
for $f$. This equation has the exact solution
\begin{equation}\label{f}
f=\frac{1}{4}\,\Big(\frac{\rho}{\rho_{\max}}
\Big)^{-\frac{1}{4}}B_{\frac{\rho}{
\rho_{\max}}}\Big(\frac{1}{4},\frac{1}{2}\Big)\,,
\end{equation}
where $B_x(a,b)$ is the incomplete Beta function and the integration
constant is fixed such that $f(\rho\rightarrow0)=1$ in order to
respect the correspondence principle. Substituting (\ref{f}) into the
relation (\ref{Friedmann2}), it is easy to obtain the modified
Friedmann equation (\ref{LQC-Friedmann}) that is suggested by LQC.
Although the Friedmann equation is the same as that is suggested by
LQC, note that our setup differs from the LQC scenario since the
energy conservation relation (\ref{Bianchi2}) is different from the
standard one (\ref{LQC-ECR}). This is because of the fact that the
matter part is not modified in LQC setup while it gets UV
modifications in rainbow cosmology through the modified dispersion
relation \cite{DSR-DoS,DSR-Inflation} (see also section 5). For the
case of radiation dominated universe, the modification to the matter
part results in deformed equation of state parameter
\cite{DSR-DoS,DSR-Inflation}. Substituting (\ref{f}) into the
relation (\ref{Bianchi2}) gives the following expression for the
equation of state parameter
\begin{equation}\label{w}
w=\frac{4}{3}\Big(1-\frac{\rho}{\rho_{\max}}\Big)\,_2F_1
\left(\frac{3}{4},1,\frac{5}{4},\frac{\rho}{\rho_{\max}}
\right)-1\,,
\end{equation}
where $_2F_1(a,b;c;x)$ is the Hypergeometric function. The above
relation correctly reduces to the standard case at low energy
regime as $\lim_{\frac{\rho}{\rho_{\max}}\rightarrow0}w=w_0=\frac{1
}{3}$.

Substituting from (\ref{equivalence}) into the relation (\ref{f})
gives $f=\frac{B_{\lambda{E}}\left(\frac{1}{4},\frac{1}{2}\right)}{
4(\lambda{E})^{\frac{1}{4}}}$ which leads to the following modified
dispersion relation
\begin{equation}\label{MDR-LQC}
\frac{B_{\lambda{E}}^2\left(\frac{1}{4},
\frac{1}{2}\right)}{16\sqrt{\lambda{E}}}
\big(E^2-p^2\big)=m^2\,,
\end{equation}
through the general relation (\ref{MSR}). Therefore, among all
possible forms for the undetermined function $f$, the special form
(\ref{f}), leading to the modified dispersion relation (\ref{MDR-LQC}),
is selected by LQC setup. In other words, the modified dispersion
relation (\ref{MDR-LQC}) leads to the effective Friedmann equation
that is suggested by LQC through the gravity's rainbow formalism.

\section{Cosmological Implications}
In the standard setup of LQC, the geometrical part of the Einstein's
equations gets modification such that the Hubble parameter and the
size of the universe (scale factor) get maximum and minimum bounds
respectively. For the case of early radiation dominated universe,
existence of these {\it geometrical} bounds immediately imply maximum
energy and entropy density bounds for the {\it matter sector}. The
matter sector is described by the standard statistical mechanics which
does not predict these bounds. More precisely, the energy and entropy
density bounds can be accommodated by the standard thermostatistics,
but they do not naturally emerge in this framework. In Ref.
\cite{DSR-DoS2}, it is shown that the energy and entropy densities get
maximum bounds when the DSR setup is applied to the statistical systems.
It is then claimed that the statistical mechanics in DSR framework is
consistent with the modified geometry suggested by LQC. This
consistency will become clear in this paper since in our
setup the geometrical modifications are exactly the same as the LQC
scenario and furthermore the statistical mechanics setup naturally
gets minimal length modifications through the modified dispersion
relation (\ref{MDR-LQC}) (see the next section). In this section, we
study the cosmological implications of our setup and the associated
statistical mechanics will be considered in the next section.

\begin{figure}
\centering
\begin{subfigure}{.49\textwidth}
  \centering
  \includegraphics[width=.9\linewidth]{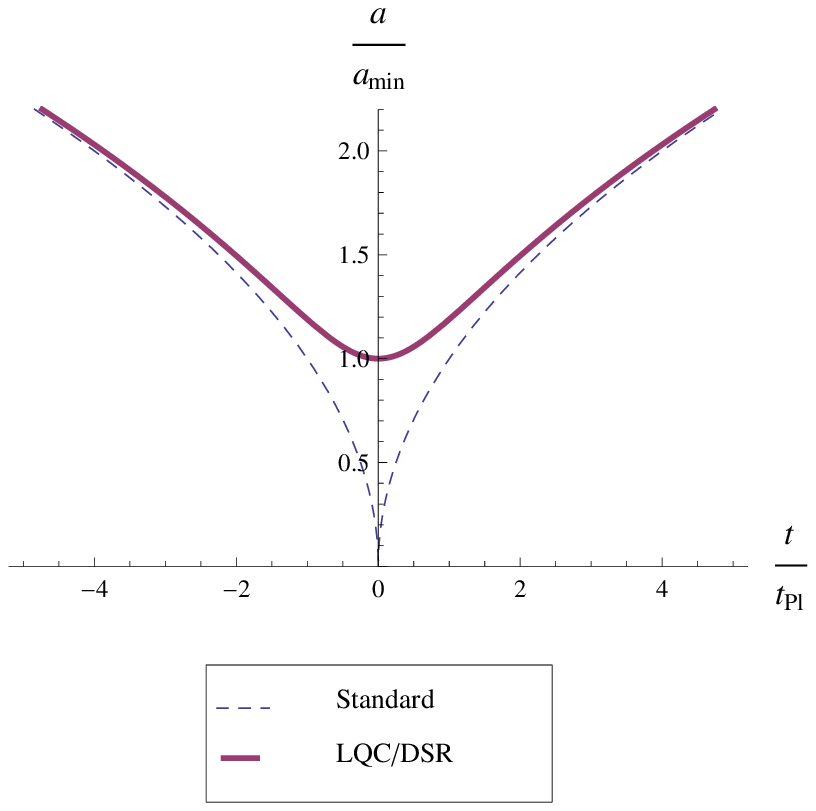}
  \caption{Scale factor versus the cosmic time. The dashed lines represent
  two separate solutions of the standard radiation
  dominated universe while the solid curve shows the unique
  bouncing trajectory in our setup.}
  \label{fig:sub1}
\end{subfigure}%
\hfill%
\begin{subfigure}{.49\textwidth}
  \centering
  \includegraphics[width=.9\linewidth]{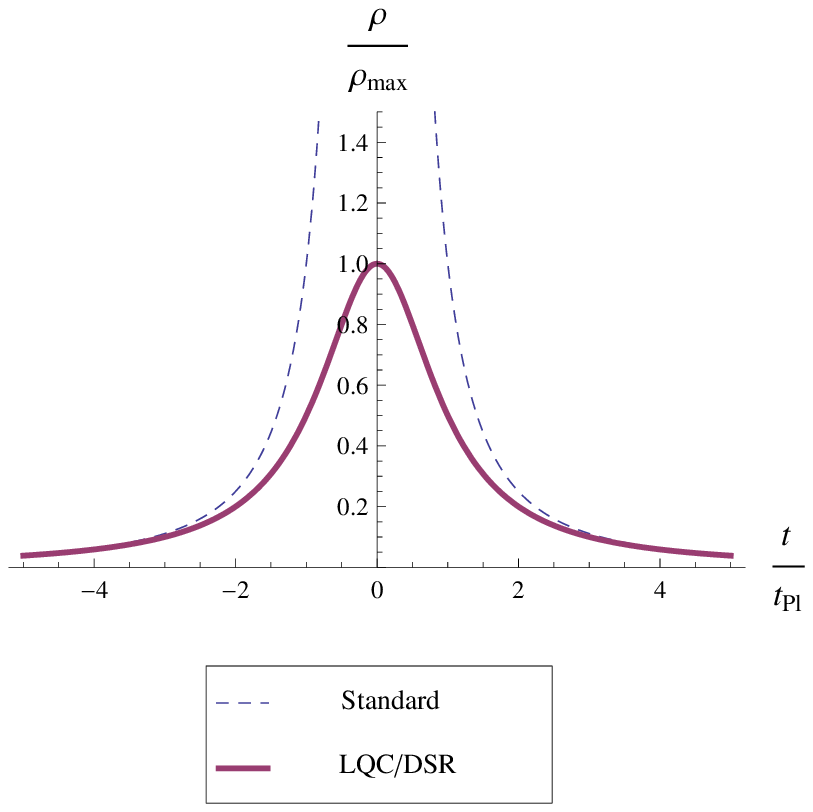}
  \caption{Energy density versus the cosmic time.
  The dashed lines correspond to two separate
  solutions for the energy density in standard radiation
  dominated universe while the solid curve shows the
  corresponding unique solution in our setup.}
  \label{fig:sub2}
\end{subfigure}
\caption{The figures are plotted for $t_{_{\rm Pl}}=
\frac{2}{\sqrt{\alpha_0}\gamma}$ and $t_b=0$.}
\label{fig:1}
\end{figure}

The conservation relation (\ref{LQC-ECR}) gives the well-known
solution $\rho=\rho_0a^{-4}$ for the energy density of the radiation
dominated universe. Although the form of the dependence of the
energy density to the scale factor is the same as the standard
radiation dominated universe, it is important to note that the
energy density has maximum in this setup as $\rho_{\max}=\rho_0
a_{\min}^{-4}$ where $a_{\min}$ is the minimum size of the
universe. Substituting this into the Friedmann equation
(\ref{LQC-Friedmann}) and then integrating the resultant equation
gives the following solution for the scale factor
\begin{equation}\label{sf}
a=a_{\min}\left(1+\frac{4}{\alpha_0\gamma^2}\left(\frac{
t-t_b}{t_{_{\rm Pl}}}\right)^2\right)^{\frac{1}{4}}\,,
\end{equation}
where the integration constant is fixed such that at $t=t_b$,
the scale factor approaches its minimum value, $a=a_{\min}$. In Eq. (\ref{sf})
we have also used the relation (\ref{rho-c}) and the
fact that $G_0=t_{_{\rm Pl}}^2$ in our units. The energy density
$\rho=\rho_0a^{-4}$ in terms of the cosmic time $t$ then can be easily
obtained as
\begin{equation}\label{rho-t}
\rho=\rho_{\max}\left(1+\frac{4}{\alpha_0\gamma^2}\left(\frac{
t-t_b}{t_{_{\rm Pl}}}\right)^2\right)^{-1}\,.
\end{equation}
It is interesting to note that the solutions (\ref{sf}) and (\ref{rho-t}) are
well-known results in LQC. The scale factor versus the cosmic time is plotted in
figure (\ref{fig:sub1}). The dashed lines in this figures show two standard
separate solutions which are disconnected by a classically
forbidden region. The left hand side dashed line corresponds to the contracting
universe ending in a future singularity and the right hand side dashed line
represents the expanding universe that starts from the past singularity.
The solid curves however, shows the solution of (\ref{sf}) which shows that the
universe in our setup (much similar to the LQC setup) bounces from
a contracting phase $t<0$ to a re-expanding epoch $t>0$. The energy
density versus the time is also plotted in figure (\ref{fig:sub2})
where again the dashed lines represent the standard classical
solutions corresponding to the contracting (the left hand side dashed line) and
expanding (the right hand side dashed line) universes. These solutions are
disconnected by a classically forbidden region and also both of them
diverge at the big bang singularity in the standard cosmology. This
classically forbidden region is replaced with a bounce in our setup
and the energy density (\ref{rho-t}) (the solid curve) increases with
time for $t<0$ and approaches its maximum value $\rho_{\max}=
\rho_0a_{\min}^{-4}$ at $t=t_b=0$, then decreases for $t>0$ and
finally coincides with its standard counterpart at $t\gg{
t_{_{\rm Pl}}}$.

Substituting from (\ref{rho-c}) for $\rho_{\max}$, the minimum size
of the universe will be
\begin{equation}\label{minimum-LQC}
a_{\min}=\left(\frac{8\pi\alpha_0\gamma^2}{3}\right)^{
\frac{1}{4}}\,l_{_{\rm Pl}}\,.
\end{equation}
This result shows that the big bang singularity problem is resolved in
our setup so that the singularity replaces with a bounce. This is
because of the fact that the geometry of our setup is dictated by LQC
scenario \cite{LQC-BB}.

\begin{figure}
\centering
\begin{subfigure}{.49\textwidth}
  \centering
  \includegraphics[width=.9\linewidth]{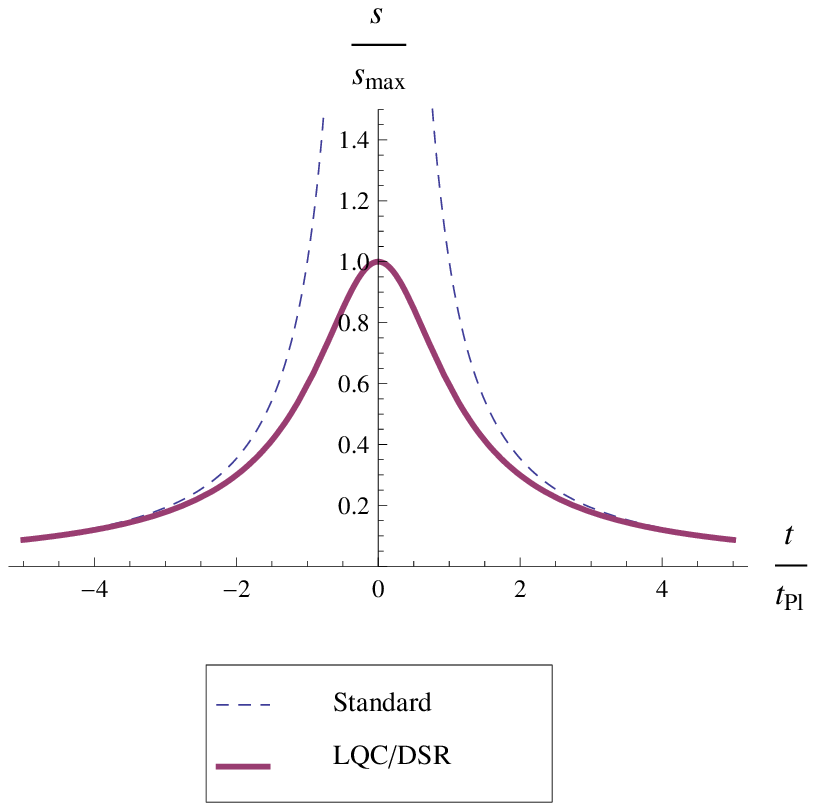}
  \caption{Entropy density versus the cosmic time.
  Dashed lines represent two separate solutions for the entropy density of the
  standard radiation dominated universe. The
  solid curve shows the corresponding unique
  solution in our setup.}
  \label{fig:sub3}
\end{subfigure}%
\hfill%
\begin{subfigure}{.49\textwidth}
  \centering
  \includegraphics[width=.9\linewidth]{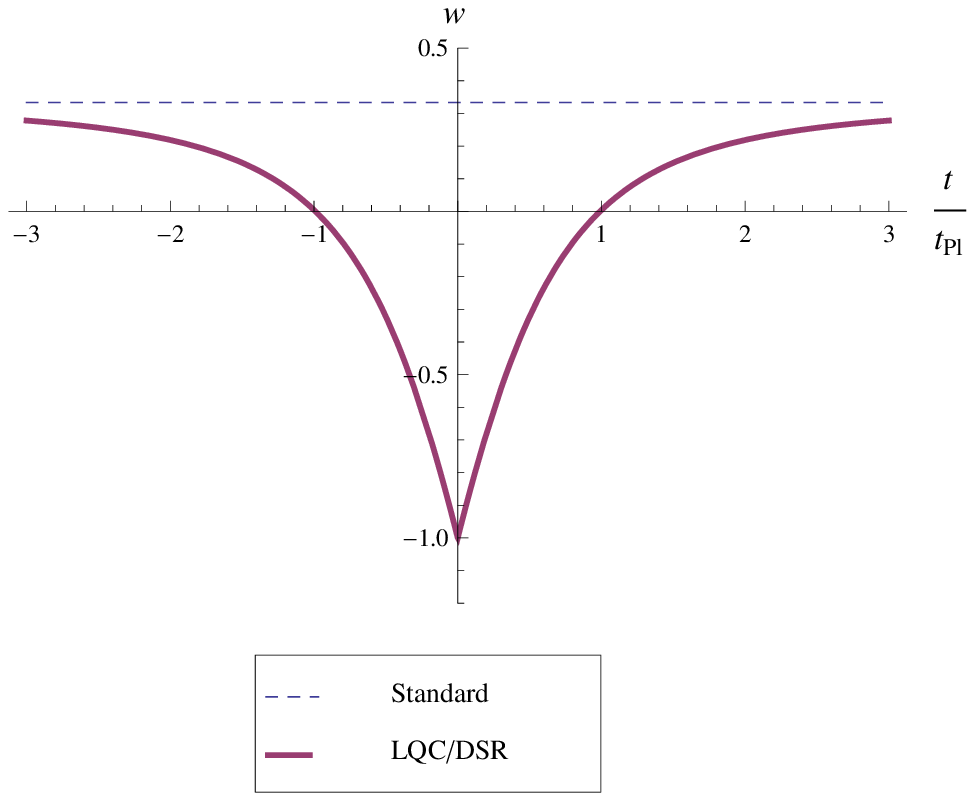}
  \caption{Equation of state parameter versus the cosmic
  time. The equation of state parameter
  (the solid curve) becomes a function of time in
  our setup while it is constant in standard
  radiation dominated universe and also in LQC
  setup (the dashed line).}
  \label{fig:sub4}
\end{subfigure}
\caption{The figures are plotted for $t_{_{\rm Pl}}=
\frac{2}{\sqrt{\alpha_0}\gamma}$ and $t_b=0$.}
\label{fig:2}
\end{figure}

Moreover, the existence of minimum size for the universe implies
maximum bound for the entropy density through the adiabatic
condition as
\begin{equation}\label{adiabatic}
sa^3=S_{\rm tot}=\mbox{constant}\,,
\end{equation}
where $S_{\rm tot}$ is the total entropy density of the universe.
Substituting the scale factor from (\ref{sf}), the time evolution
of the entropy density is then given by
\begin{equation}\label{entropy}
s=s_{\max}\left(1+\frac{4}{\alpha_0\gamma^2}\left(\frac{
t-t_b}{t_{_{\rm Pl}}}\right)^2\right)^{-\frac{3}{4}}\,,
\end{equation}
where $s_{\max}=S_{\rm tot}a_{\min}^{-3}$ is the maximum value for
the entropy density. The entropy density (\ref{entropy}) versus the cosmic
time is plotted in figure (\ref{fig:sub3}). The dashed lines at the
regions $t<0$ and $t>0$ show the two separate standard solutions for
the entropy density which are disconnected by a classically forbidden
region. Both of these solutions diverge at the big bang singularity
(the point $t=0$ in figure (\ref{fig:sub3})). Interestingly, in our
setup, the classically forbidden region replaces with a bounce such
that the entropy density (\ref{entropy}) increases with time for
$t<0$ until it approaches the maximum value $s_{\max}=S_{\rm tot}
a_{\min}^{-3}$ at $t=t_b=0$ and then decreases with time and
reduces to its standard counterpart for $t\gg{t_{_{\rm Pl}}}$.

While the equation of state remains unchanged in LQC setup (since
only the geometry gets modified), it significantly modifies at
UV regime in our setup through the deformed density of states. This
is the crucial difference of our model with LQC at the cosmological
setup. Substituting (\ref{rho-t}) into (\ref{w}), we can easily
obtain the time evolution of the equation of state parameter which is
plotted in figure (\ref{fig:sub4}). The negative values near the
bounce $t=t_b=0$ signal repulsive force which is responsible for the
singularity resolution in our setup. This repulsive force could
naturally solve the horizon problem \cite{rainbow}. Furthermore, such a
unstable accelerating phase has been interpreted as an inflationary phase
in radiation dominated universe in Ref. \cite{DSR-Inflation}.

The spacetime singularity resolution that is emerged in our setup is
a common feature of LQC scenario. More precisely, except the
time-dependence of equation of state parameter (which is shown in
figure (\ref{fig:sub4})), all the results of this section are the same
as those suggested by LQC. But, the advantage of our setup is that
we can explore the statistical origin of the entropy and energy
density bounds (see figures (\ref{fig:sub2}) and (\ref{fig:sub3}))
while one cannot do so in LQC scenario. This is because of that fact
that both of the geometrical and matter parts of Einstein's equations
get modified in gravity's rainbow formalism while in LQC setup
only the geometry is modified.

\section{Entropy and Energy Density Bounds}
In this section, we would like to study the thermostatistical properties
of the early radiation dominated universe in our model defined by the
modified dispersion relation (\ref{MDR-LQC}). The key quantity is the
number of states density which determines the total number of accessible
microstates for the system under consideration. For the case of massless
particles with which we are interested in early universe thermodynamics,
the modified dispersion relation (\ref{MDR-LQC}) gives $E=p$ and
therefore the form of number of states density remains unchanged in our
setup (see Refs. \cite{DSR-DoS2,DSR-Inflation,DSR-DoS} for details).
But, it is important to note that there is UV cutoff $E<\lambda^{-1}$
(or equivalently $p<\lambda^{-1}$ since $E=p$) in our setup and thus the
modifications to the thermodynamical quantities arise at high energy
regime $E\sim{\lambda^{-1}}\sim{E_{_{\rm Pl}}}$. The number of
states density with energy in the interval between $E$ and $E+dE$ is
given by \cite{DSR-DoS}
\begin{equation}\label{DoS}
d\Gamma=\frac{4\pi{V}g_{\star}}{h^3}E^2dE\,,
\end{equation}
where $g_{\star}$ is the effective number of internal degrees of
freedom and $h=2\pi$ in our units. The internal energy and the number
of particles in local region with spatial volume $V$ are defined as
\begin{equation}\label{IE}
U=\int{n(E)}E\,d\Gamma\,,
\end{equation}
\begin{equation}\label{N}
N=\int{n(E)}\,d\Gamma\,,
\end{equation}
where $n(E)=\left(\exp(E/T)\mp1\right)^{-1}$ is the ensemble density
and $(-)$ and $(+)$ signs are corresponding to the bosons and fermions
respectively. The form of the ensemble density may be different at
very high energy regime $E\sim{E_{_{\rm Pl}}}$ such that the standard
definitions of bosons and fermions are no longer to be applicable.
But, fortunately, following the method that is introduced in Ref.
\cite{DSR-DoS2} we can address the main statistical properties of the
universe without any attribution to ensemble density. The total
number of microstates can be obtained by directly integrating the
number of states density (\ref{DoS}) as
\begin{equation}\label{TNoM}
\Gamma=\int{d\Gamma}=\frac{Vg_{\star}}{2\pi^2}\int_{0}^{
\lambda^{-1}}E^2dE=\frac{g_{\star}}{6\pi^2}\left(\frac{V
}{\lambda^3}\right)\,,
\end{equation}
which interestingly is turned out to be finite. Note that the total
number of microstates in the standard early universe statistical
mechanics diverges since there is no upper bound for the
kinematical energy and therefore the system could access more
and more microstates with high and higher energies. The universe then
goes to big bang singularity at $E\rightarrow\infty$ in the standard
model of cosmology. The existence of the upper bound $E<\lambda^{-1}$
however prevents the big bang singularity from the statistical point
of view such that the total number of potentially accessible
microstates for the universe becomes finite. Having finite total
number of microstates (\ref{TNoM}) immediately implies maximum bounds
for the energy and entropy densities from the statistical point of
view \cite{DSR-DoS2}.

The entropy is directly determined by the number of microstates and
therefore the emergence of finite total number of microstates
(\ref{TNoM}) immediately leads to the following entropy density
bound
\begin{equation}\label{S}
s_{\max}=\ln{\Gamma}=\ln\left(\frac{V}{\lambda^3}\right)
+\ln\left(\frac{g_{\star}}{6\pi^2}\right)\,,
\end{equation}
for the early radiation dominated universe in our setup. From the
relations (\ref{TNoM}) and (\ref{S}) it is clear that, at high
energy regime, the total number of microstates in local region with
spatial volume $V$ is completely determined by the ratio $\frac{V}{
\lambda^3}\sim\frac{V}{l_{_{\rm Pl}}^3}$. This result shows that one
bit of information is determined by the fundamental volume
$l_{_{\rm Pl}}^3$ in short distance regime.

Also, from the definition (\ref{N}), it is clear that the total number
of particles in a local region $V$ becomes finite as $N_{\max}=\int{
d\Gamma}=\Gamma$ where we have used the relation (\ref{TNoM}). The
associated total internal energy is defined by $U_{\max}=N_{\max}
\times\frac{\int{E}d\Gamma}{\Gamma}=\int{E}d\Gamma$ which gives
\begin{equation}\label{rho-max}
\rho_{\max}=\frac{g_{\star}}{2\pi^2}\int_{0}^{\lambda^{
-1}}E^3dE=\frac{g_{\star}}{8\pi^2}\lambda^{-4}\,,
\end{equation}
where $\rho_{\max}=U_{\max}/V$ is the maximum accessible energy
density for the universe. Thus, the existence of the upper bound for
$E\leq\lambda^{-1}$ for the kinematical energy leads to the finite
total number of microstates for the universe which immediately
implies the maximum entropy and energy density bounds (\ref{S}) and
(\ref{rho-max}). These results explain the statistical origin of the
entropy and energy density bounds which are shown in figures
(\ref{fig:sub2}) and (\ref{fig:sub3}).

We are now adequately equipped to fix explicitly the equivalence
relation (\ref{equivalence}) that we have assumed in
section 3. From a thermodynamical point of view, the energy of the
universe is given by the internal energy $U$ (or equivalently by the
internal energy density $\rho=U/V$). On the other hand, from the kinematical point of view
the energy is determined by the kinematical energy $E$. For the case
of radiation dominated universe, the internal energy
is directly related to the kinematical energy $U\sim{E}$. More
precisely, the internal energy $U$ is the statistical average of
the kinematical energy $E$ through the well-known definition (\ref{IE}).
In our setup, there is an upper bound $E\leq\lambda^{-1}$ for the
kinematical energy that nontrivially implies an upper bound
(\ref{rho-max}) for the energy density (see also Ref.
\cite{DSR-DoS2} where it is shown that the existence of an upper
bound on kinematical energy does not necessarily imply an upper
bound on the internal energy). The energy density should then
approach its maximum value (\ref{rho-max}) when the kinematical
energy approaches the maximum value $E_{\max}=\lambda^{-1}$.
Therefore, in our setup, the critical energy density $\rho_{\max}$
(that is obtained in LQC setup) will be exactly equal to the maximum
energy density $\rho_{\max}=U_{\max}/V$ which is obtained through
statistical considerations. From (\ref{rho-c}) and (\ref{rho-max}),
the observer-independent length scale $\lambda$ becomes fixed completely
in terms of Barbero-Immirzi parameter $\gamma$ and effective number
of degrees of freedom $g_{\star}$ as
\begin{equation}\label{minimal-length}
\lambda=\left(\frac{\alpha_0\gamma^2g_{\star}}{3\pi}\right)^{
\frac{1}{4}}\,l_{_{\rm Pl}}\,.
\end{equation}
Substituting (\ref{minimal-length}) into (\ref{minimum-LQC}), gives
the relation between the minimum size of the universe and the
observer-independent length scale of DSR model (\ref{MDR-LQC}) as
\begin{equation}\label{minimum-SF}
a_{\min}=\left(\frac{8\pi^2\rho_0}{g_{\star}}\right)^{\frac{
1}{4}}\,\lambda\,.
\end{equation}
The result (\ref{minimal-length}) is interesting since it shows the
explicit relation between the observer-independent length scale of
DSR theories and Planck length as quantum of length that is suggested by LQC
scenario. This relation also explicitly confirms that the DSR
theories can be considered as the flat limit of loop quantum gravity
\cite{DSR-FL}. Note that the minimum size of the universe
(\ref{minimum-LQC}) is completely fixed in LQC scenario since
$\gamma=0.2375$ (through the black hole entropy calculation \cite{BI})
and $\alpha_0=4\sqrt{3}\pi\gamma=5.166$. But, the minimum
observer-independent length scale $\lambda$ turned out to be directly
related to the total number of particles that contribute to the
energy content of the radiation dominated universe through the
effective number of degrees of freedom $g_{\star}$. Bounds on
$\lambda$ and therefore on $g_{\star}$ can be obtained through the
methods that are introduced in Refs. \cite{QGExpriment}.

\section{Summary and Conclusions}

Existence of a minimum measurable length scale is suggested by quantum
gravity candidates such as loop quantum gravity and string theory.
It is therefore widely believed that, at the flat limit when purely
gravitational and quantum mechanical effects are negligible, the
ultimate quantum theory of gravity would reduce to a deformed (doubly)
special relativity (DSR) which supports the existence of minimum
observer-independent length scale. Gravity's rainbow (or doubly
general relativity) is a simple generalization of DSR theories to
include gravity in the semiclassical regime. In cosmological setup,
different DSR models lead to the different Friedmann equations which
include the effects of minimal observer-independent length scale. On
the other hand, inspired by loop quantum gravity and string theory,
the modified Friedmann equations are investigated in loop quantum
cosmology (LQC) and string cosmology scenarios which also include
the minimal length effects. In this paper, we have introduced a method
by which one can fix the one-parameter family of Friedmann equations
in rainbow cosmology by means of the deformed Friedmann equations that
are suggested by the mentioned quantum gravity candidates. We applied
the setup to the case of unique modified Friedmann equation suggested
by LQC scenario. In this respect, among all possible DSR models, the
DSR model defined by the modified dispersion relation (\ref{MDR-LQC})
is selected by LQC. Moreover, this phenomenological derivation of LQC
equations allowed us to explore the statistical origin of the energy
and entropy density bounds that are arisen from the purely geometrical
considerations in LQC scenario. This is because of the fact that while
the statistical mechanics formalism remains unchanged in LQC scenario,
it naturally gets quantum gravity modifications in our setup. We
found that the total number of accessible microstates for the universe
is finite which immediately implies the maximum bounds for the energy
and entropy densities from the statistical point of view. The
statistical considerations also completely fixed the relation between
the quantum of spacetime in LQC setup in one side and the
observer-independent length scale in DSR theories as
(\ref{minimal-length}) on the other side. These results explicitly
show the relevance of the DSR theories for the flat limit of loop
quantum gravity.

\end{document}